\documentclass[a4paper,11pt]{article}
\usepackage{amsmath,amssymb,color,comment,cancel}
\usepackage[normalem]{ulem}

\usepackage{jheppub} 

\usepackage[T1]{fontenc} 

\title{
Negative Mode Problem of False Vacuum Decay Revisited
}

\author[a]{Ryusuke Jinno}
\author[a]{and Ryosuke Sato}

\affiliation[a\,]{DESY, Notkestra\ss e 85, D-22607 Hamburg, Germany}

\emailAdd{ryusuke.jinno@desy.de}
\emailAdd{ryosuke.sato@desy.de}

\abstract{
We study negative modes around the Coleman-de Luccia bounce solution.
The conditions for the appearance of an infinite number of the negative modes do not coincide between Lagrangian and Hamiltonian formulations in the literature, and we discuss the origin of this difference in detail.
We show how different choices of the variable for the fluctuation around the bounce solution affect the negative mode condition in the Hamiltonian approach, and point out that there exists a choice which gives the same negative mode condition as the Lagrangian approach.
}

\begin{document} 
\rightline{DESY 20-171}
\maketitle
\flushbottom

\section{Introduction}

False vacua appear in a wide class of models in particle physics.
Such vacua decay into the true vacuum via quantum tunneling within a finite lifetime.
The lifetime of the false vacua gives important implications for model building and cosmology.

Coleman and Callan proposed calculating the decay width of the false vacua by utilizing the Euclidean path integral \cite{Coleman:1977py, Callan:1977pt}.
In this formalism, the ``imaginary part'' of the energy eigenvalue of the vacuum is interpreted as the decay width.
The dominant contribution for the decay width in this path integral comes from the bounce solution,
which gives the least action among non-trivial solutions of the equation of motion for the scalar fields.
Since the bounce solution is a saddle point of the action,
there exists a mode of fluctuation which makes the action smaller.
Such a mode is called a \textit{negative mode}.
The existence of one and only one negative mode is proved in Ref.~\cite{Coleman:1987rm} in a wide class of models.
This one negative mode plays an essential role in giving the imaginary part of the path integral.

Coleman and De Luccia (CDL) \cite{Coleman:1980aw} extended this study to the case with gravity.
They claimed that a saddle point configuration of the Euclidean gravity action should give the dominant contribution to the decay width.
However, the CDL bounce solution has a puzzling property.
In the CDL case, the existence of only one negative mode is not guaranteed,
and an infinite number of negative modes can exist in some situation \cite{Lavrelashvili:1985vn}.
This problem has been known as the negative mode problem and it is not clear how we should interpret the existence of those modes.
Despite quite a few studies \cite{Tanaka:1992zw, Tanaka:1999pj, Khvedelidze:2000cp, Lavrelashvili:1999sr, Gratton:2000fj, Dunne:2006bt, Lee:2014uza, Koehn:2015hga, Bramberger:2019mkv},
there is no conclusive answer to this problem yet.
(For other problems in the Euclidean path integral with gravity, see, \textit{e.g.}, Ref.~\cite{Hebecker:2018ofv}.
In the context of axion wormholes, see, \textit{e.g.},  Ref.~\cite{Hertog:2018kbz}.)

There has been a confusion about the condition under which those infinite number of negative modes appear.
One formulation to derive the condition is based on the expansion of the action around the bounce solution
by Lavrelashvili, Rubakov, and Tinyakov (LRT) \cite{Lavrelashvili:1985vn} and Lee and Weinberg (LW) \cite{Lee:2014uza}
(see also approach I in Lavrelashvili (L-I) \cite{Lavrelashvili:1999sr}).
In addition to this, there is other formulations based on the Hamiltonian of the fluctuation
by Khvedelidze, Lavrelashvili, and Tanaka (KLT) \cite{Khvedelidze:2000cp}
(see also Gratton and Turok (GT) \cite{Gratton:2000fj} and approach III in Lavrelashvili (L-III) \cite{Lavrelashvili:1999sr}),
and by Garriga, Montes, Sasaki, and Tanaka (GMST) \cite{Garriga_1998}
(see also approach II in Lavrelashvili (L-II) \cite{Lavrelashvili:1999sr}),
and these formulations give different conditions from Refs.~\cite{Lavrelashvili:1985vn, Lee:2014uza}.
Thus, the problems on the false vacuum decay with gravity is twofold: one is that
an appropriate treatment of infinite number of negative modes is not known (this is exactly the negative mode problem),
and the other is that different conditions on the appearance of negative modes are presented among the literature.
The ultimate goal is solving the former problem and this automatically includes the solution for the latter problem.
Before going to this stage, we tackle on the latter problem, \textit{i.e.},
the goal of this paper is to clarify the origin of the different conditions in the Lagrangian and Hamiltonian formulations.
As we will see, we can mix a field and its conjugate momentum by a canonical coordinate transformation in the Hamiltonian formulation,
and this is the reason why the literature give different negative mode conditions.

The rest of this paper is organized as follows.
In Sec.~\ref{sec:lagrangian formulation}, we derive the negative mode condition by the Lagrangian formulation.
In Sec.~\ref{sec:hamiltonian formulation}, we derive the condition by the Hamiltonian formulation.
In Sec.~\ref{sec:summary and discussion}, we give summary and discussions.
In Appendix \ref{sec:comparison}, we discuss the negative mode conditions in Refs.~\cite{Lavrelashvili:1985vn, Lee:2014uza, Khvedelidze:2000cp, Gratton:2000fj, Lavrelashvili:1999sr} in detail.

\section{Lagrangian formulation} \label{sec:lagrangian formulation}

We follow the notation in Ref.~\cite{Lavrelashvili:1999sr}.
The Euclidean action for a real scalar field with gravity is given as
\begin{align}
S = \int d^4 x \sqrt{g} \left[ -\frac{1}{2\kappa} R + \frac{1}{2} (\partial_\mu \phi) (\partial^\mu\phi) + V(\phi) \right], \label{eq:action}
\end{align}
where $\kappa = 8 \pi G$.
In order to discuss the $O(4)$ symmetric background\footnote{
A proof of $O(4)$ symmetry of the bounce solution with gravity is not known though it is widely believed.
For the case without gravity, see Refs.~\cite{Coleman:1977th, lopes1996radial, byeon2009symmetry, Blum:2016ipp}.}
and fluctuations around it,
we parametrize the metric and the scalar field as
\begin{align}
ds^2 &= a(\tau)^2 [ (1+2A(\tau)) d\tau^2 + \gamma_{ij} ( 1-2\Psi(\tau)) dx^i dx^j ], \label{eq:metric} \\
\phi &= \varphi(\tau) + \Phi(\tau). \label{eq:scalar}
\end{align}
Here $A$, $\Psi$, and $\Phi$ are perturbations around the background, and $\gamma_{ij}$ is the three dimensional metric with a constant spatial curvature.
We assume the latter to be positive definite in this paper.
For the background field $a$ and $\varphi$, we obtain the following field equations:
\begin{align}
{\cal H}^2 - {\cal H}' - {\cal K} &= \frac{\kappa}{2}\varphi'^2, \\
2{\cal H}' + {\cal H}^2 - {\cal K} &= -\frac{\kappa}{2}(\varphi'^2 + 2a^2 V), \\
\varphi'' + 2{\cal H}\varphi' - a^2 \frac{\partial V}{\partial\varphi} &= 0.
\end{align}
Here $X' \equiv dX/d\tau$ and ${\cal H} \equiv a'/a$.
The curvature ${\cal K} = +1,~0,-1$ are for closed, flat, and open universes, respectively (with an appropriate unit).
We take ${\cal K}=1$ for the perturbation around the CDL bounce solution.
The case with ${\cal K} = 0$ tells us about the zero-mode of inflaton in a flat universe. 
For details, see Appendix \ref{sec:inflation}.

Let us expand the Lagrangian around the background field as
\begin{align}
{\cal L} = {\cal L}^{(0)} + {\cal L}^{(2)} + \cdots,
\end{align}
where ${\cal L}^{(0)}$ does not include $A$, $\Psi$, and $\Phi$, and ${\cal L}^{(2)}$ consists of quadratic term of them
\begin{align}
{\cal L}^{(2)} = \frac{a^2 \sqrt{\gamma}}{2\kappa} \Biggl[
&- 6\Psi'^2 + 6{\cal K}\Psi^2 + \kappa \left(\Phi'^2 + a^2 \frac{\partial^2 V}{\partial\varphi^2}\Phi^2 + 6\varphi' \Psi' \Phi \right) \nonumber\\
&- \left(
12 {\cal H} \Psi' + 12 {\cal K} \Psi + 2\kappa \varphi' \Phi' - 2\kappa a^2 \frac{\partial V}{\partial\varphi} \Phi 
\right) A 
- 2({\cal H}' + 2{\cal H}^2 + {\cal K}) A^2
\Biggr].
\label{eq:quadratic}
\end{align}
The action Eq.~(\ref{eq:action}) is invariant under a general coordinate transformation.
With the parametrization as Eqs.~(\ref{eq:metric}) and (\ref{eq:scalar}), the action is invariant under a reparametrization of $\tau$: $\tau \to \tau'(\tau)$. 
This invariance is inherited as a gauge invariance of ${\cal L}^{(2)}$.
Infinitesimal gauge transformations for $A$, $\Psi$, and $\Phi$ are given as
\begin{align}
\delta A = \frac{\epsilon'}{a}, \qquad
\delta\Psi = -\frac{\epsilon}{a} {\cal H}, \qquad
\delta\Phi = \frac{\epsilon}{a}\varphi'.
\label{eq:gauge_transformation}
\end{align}
Indeed, we can write down the action (\ref{eq:quadratic}) in a gauge-invariant form:
\begin{align}
{\cal L}^{(2)} 
&= 
a^2 \sqrt{\gamma}
\Biggl[
\frac{\varphi'^2}{2{\cal H}^2} {\cal R}'^2
- \frac{1}{2{\cal H}^2}
\left(
\frac{6{\cal K}^2}{\kappa} + 3{\cal K} \varphi'^2
\right)
{\cal R}^2
- \frac{\varphi'^2}{{\cal H}} {\cal R}' {\cal A}
- \frac{6{\cal K}}{\kappa} {\cal R} {\cal A}
- \frac{3 {\cal H}^2}{\kappa} {\cal A}^2 + \frac{1}{2} {\cal A}^2 \varphi'^2
\Biggr].
\end{align}
Here ${\cal R}$ and ${\cal A}$ are gauge-invariant combinations:
\begin{align}
{\cal R}
&= 
\Psi + \frac{{\cal H}}{\varphi'} \Phi,
\\
{\cal A}
&= 
A + \frac{1}{{\cal H}} \Psi' + \left( \frac{{\cal H}'}{{\cal H} \varphi'} - \frac{{\cal H}}{\varphi'} \right) \Phi.
\end{align}
Integrating ${\cal A}$ out and performing integration by parts, we get
\begin{align}
{\cal L}^{(2)}
=
a^2 \sqrt{\gamma}
\Biggl[
\frac{3 \varphi'^2}{6{\cal H}^2 - \kappa \varphi'^2} {\cal R}'^2
+ 
\frac{6 {\cal K}}{(6{\cal H}^2 - \kappa \varphi'^2)^2}
\left(
6 {\cal H}^2 \varphi'^2 - 6 {\cal K} \varphi'^2 - \kappa \varphi'^4 - 6 {\cal H} \varphi' a^2 \frac{\partial V}{\partial \varphi}
\right)
{\cal R}^2
\Biggr].
\label{eq:L_gaugeinv}
\end{align}
Therefore, we can directly read off the coefficient of the kinetic term of ${\cal R}$ to obtain the negative mode condition:
\begin{align}
Q &\equiv 
{\cal H}^2 - \frac{\kappa}{6} \varphi'^2.
\label{eq:definition Q}
\end{align}
The coefficient of the kinetic term of ${\cal R}$ becomes negative if $Q<0$.
This result is consistent with LRT \cite{Lavrelashvili:1985vn}, LW \cite{Lee:2014uza}, and L-I \cite{Lavrelashvili:1999sr}.

\section{Hamiltonian formulation}\label{sec:hamiltonian formulation}

In this section, we discuss the Hamiltonian formulation for the perturbations around the background field.
Let us start from the Lagrangian given in Eq.~(\ref{eq:quadratic}).
Conjugate momenta for $\Psi$, $\Phi$, and $A$ are defined as
\begin{align}
\Pi_A &\equiv \frac{\partial {\cal L}^{(2)}}{\partial A'} = 0, \\
\Pi_{\Psi} &\equiv \frac{\partial {\cal L}^{(2)}}{\partial \Psi'} 
= \frac{6 a^2\sqrt{\gamma}}{\kappa}\left( -\Psi' + \frac{\kappa}{2}\varphi' \Phi - {\cal H} A \right), \\
\Pi_{\Phi} &\equiv \frac{\partial {\cal L}^{(2)}}{\partial \Phi'} 
= a^2\sqrt{\gamma} \left( \Phi' - \varphi' A \right),
\end{align}
Thus, the Lagrangian ${\cal L}^{(2)}$ is singular and we have a primary constraint
\begin{align}
C_1 \equiv \Pi_A = 0.
\end{align}
The total Hamiltonian \cite{Dirac} is given as
\begin{align}
H_T =& H_C + v_1(\tau) C_1, \label{eq:HT}\\
H_C =&
-\frac{\kappa}{12a^2\sqrt{\gamma}} \Pi_\Psi^2 + \frac{1}{2a^2\sqrt{\gamma}} \Pi_\Phi^2 + \frac{1}{2}\kappa \varphi' \Pi_\Psi \Phi
- a^2 \sqrt{\gamma}\left[ \frac{3{\cal K}}{\kappa} \Psi^2 + \frac{1}{2}\left( a^2 \frac{\partial^2 V}{\partial\varphi^2} + \frac{3}{2}\kappa \varphi'^2 \right) \Phi^2 \right] \nonumber\\
&+ A\left[ \varphi' \Pi_\Phi - {\cal H} \Pi_\Psi + a^2 \sqrt{\gamma}\left( \left( 3{\cal H}\varphi' - a^2 \frac{\partial V}{\partial\varphi} \right) \Phi + \frac{6{\cal K}}{\kappa}\Psi \right) \right],
\end{align}
where $v_1$ is an arbitrary function of $\tau$.
An explicit form of $v_1$ can be determined by a gauge fixing.
The time evolution of an arbitrary variable $f$ is given as
\begin{align}
f' = \frac{\partial f}{\partial \tau} + [f,H_T],
\label{eq:Poisson}
\end{align}
where the Poisson bracket is defined as
\begin{align}
[f,g] \equiv \sum_{q=A,\Psi,\Phi}
\left(\frac{\partial f}{\partial q} \frac{\partial g}{\partial \Pi_q}
- \frac{\partial f}{\partial \Pi_q} \frac{\partial g}{\partial q} \right).
\end{align}
Note that the partial derivative acts only on background quantities in Eq.~(\ref{eq:Poisson}).
Since $C_1' = 0$ should be satisfied for consistency with $C_1 = 0$,
we obtain a secondary constraint as
\begin{align}
C_2 
&\equiv C_1' 
= \frac{\partial C_1}{\partial \tau} + [C_1, H_T] 
= - \varphi' \Pi_\Phi + {\cal H} \Pi_\Psi + a^2 \sqrt{\gamma} \left[ \left( - 3{\cal H}\varphi' + a^2 \frac{\partial V}{\partial\varphi} \right) \Phi - \frac{6{\cal K}}{\kappa}\Psi \right].
\label{eq:C1 time evolution}
\end{align}
The time evolution of $C_2$ is written in terms of $C_2$ itself
\begin{align}
C_2' = \frac{\partial C_2}{\partial \tau} + [C_2, H_T] = {\cal H}C_2, \label{eq:C2 time evolution}
\end{align}
and thus no tertiary constraint arises.
$C_1$ and $C_2$ are first-class constraints because $[C_i, C_j]=0$ ($i, j = 1,2$) and the matrix $[C_i, C_j]$ is not invertible.

As we have seen, the Lagrangian ${\cal L}^{(2)}$ is invariant up to total derivatives under the gauge transformation (\ref{eq:gauge_transformation}).
Let us discuss this gauge invariance in the light of Hamiltonian.
Although Dirac conjectured that all of the first class constraints generate gauge symmetry \cite{Dirac}, 
not all of them are generator of gauge symmetry in reality \cite{cawley1979determination}.
The number of generators of gauge symmetry is equal to the number of primary first-class constraints~\cite{Castellani:1981us, Sugano:1982br} though the generators are in general linear combinations of primary and non-primary first-class constraints.
Indeed, in the present case we have only one degree of freedom $\epsilon$ for the gauge transformation (see Eq.~(\ref{eq:gauge_transformation})).

According to Refs.~\cite{Castellani:1981us, Sugano:1982br}, a generator of gauge symmetry $G_\epsilon$ with an arbitrary function $\epsilon(\tau)$ should satisfy
\begin{align}
\frac{\partial G_\epsilon}{\partial \tau} + [G_\epsilon,H_T] = ({\rm primary~constraints}).
\end{align}
By using Eq.~(\ref{eq:C1 time evolution}) and Eq.~(\ref{eq:C2 time evolution}),
we can determine the solution of the above equation up to overall factor:
\begin{align}
G_\epsilon = - \frac{\epsilon}{a} C_2 + \frac{\epsilon'}{a}C_1.
\end{align}
By using this generator, we obtain the gauge transformation for perturbations as
\begin{align}
[\Psi,G_\epsilon] &= -\frac{\epsilon}{a} {\cal H}, \\
[\Phi,G_\epsilon] &= \frac{\epsilon}{a}\varphi',\\
[A,G_\epsilon] &= \frac{\epsilon'}{a}, \\
[\Pi_\Psi,G_\epsilon] &= - \epsilon a\sqrt{\gamma} \frac{6 {\cal K}}{\kappa},\\
[\Pi_\Phi,G_\epsilon] &= \epsilon a \sqrt{\gamma} \left( a^2 \frac{\partial V}{\partial\varphi} - 3 {\cal H} \varphi' \right) = \epsilon a\sqrt{\gamma} (\varphi'' - \varphi' {\cal H}), \\
[\Pi_A,G_\epsilon] &= 0.
\end{align}
The above transformation law is consistent with Eq.~(\ref{eq:gauge_transformation}) and Refs.~\cite{Garriga_1998,Lavrelashvili:1999sr}.

As we will discuss later, gauge fixing is required when we use the Dirac bracket.
Let us briefly comment on the gauge fixing conditions in the literature.
Ref.~\cite{Khvedelidze:2000cp} imposes $\Pi_\Psi = 0$ and $A-\Psi = 0$,
and Ref.~\cite{Lavrelashvili:1999sr} imposes $\Pi_\Psi=0$ and $A=0$ in their approach III.
The time evolution of $\Pi_\Psi$ is $[\Pi_\Psi, H_T] = -6\sqrt{\gamma} {\cal K} a^2 (A-\Psi)/\kappa$.
Thus,
the gauge fixing condition in Ref.~\cite{Khvedelidze:2000cp} ($\Pi_\Psi=0$ and $A-\Psi=0$) is consistent,
but the condition in Ref.~\cite{Lavrelashvili:1999sr} ($\Pi_\Psi=0$ and $A=0$) is inconsistent.

\subsection{Canonical transformation approach} \label{sec:canonical transf}

So far all of the variables are gauge variant in the current coordinate.
Let us move to a new coordinate system with gauge invariant variables.
We can see there exist three independent gauge invariant linear combinations of $\Psi$, $\Phi$, $\Pi_\Psi$, and $\Pi_\Phi$.
$C_2$ is one of such linear combinations, and it is convenient to adopt constraints as coordinates as discussed in Ref.~\cite{Maskawa:1976hw}.
Another convenient gauge invariant combination is 
\begin{align}
{\cal R} &\equiv \Psi + \frac{{\cal H}}{\varphi'} \Phi. \label{eq:definition R}
\end{align}
In addition to these, ${\cal R}' = \partial{\cal R}/\partial \tau + [{\cal R},H_T]$ is also a gauge invariant variable, which is given by the original variables as
\begin{align}
{\cal R}' &= \left( \frac{3 {\cal H}^2 - {\cal K}}{\varphi'} - \frac{a^2 {\cal H}}{\varphi'^2}\frac{\partial V}{\partial\varphi} \right) \Phi  + \frac{1}{\varphi'} \frac{{\cal H}}{a^2\sqrt{\gamma}} \Pi_\Phi - \frac{\kappa}{6 a^2\sqrt{\gamma}}\Pi_\Psi. \label{eq:Rprime}
\end{align}
Let us move to a coordinate system which has ${\cal R}$ and $C_2$ as canonical coordinates. 
We define their conjugate momenta as
\begin{align}
\Pi_{\cal R} &\equiv \frac{a^2 \sqrt{\gamma} \varphi'^2}{ Q } {\cal R}', \label{eq:definition PiR}\\
\Pi_{C_2} &\equiv -\frac{1}{ Q} \left( {\cal H} \Psi + \frac{\kappa}{6} \varphi' \Phi \right) + f(\tau) C_2.
\end{align}
Here $f$ is an arbitrary function of $\tau$, and $Q$ is defined in Eq.~(\ref{eq:definition Q}).
By using the above definition, we can check ${\cal R}$, $\Pi_{\cal R}$, $C_2$, and $\Pi_{C_2}$ satisfy
\begin{align}
[{\cal R},\Pi_{\cal R}] = [C_2, \Pi_{C_2}] = 1, \\
[{\cal R}, C_2] = [\Pi_{\cal R}, C_2] =  [{\cal R},\Pi_{C_2}] = [\Pi_{\cal R},\Pi_{C_2}] = 0, \\
[{\cal R},A]= [\Pi_{\cal R},A]= [C_2,A]= [\Pi_{C_2},A]=0, \\ 
[{\cal R},\Pi_A]= [\Pi_{\cal R},\Pi_A]= [C_2,\Pi_A]= [\Pi_{C_2},\Pi_A]=0.
\end{align}
Thus, we can use $({\cal R}, C_2, A, \Pi_{\cal R}, \Pi_{C_2}, \Pi_A)$ as canonical coordinates.
Hamiltonian in the new coordinate can be derived from a (type-three) generating function $W = W(\Pi_\Psi, \Pi_\Phi, {\cal R}, C_2)$. Taking the generating function as
\begin{align}
W
&=
\frac{1}{a^2 \sqrt{\gamma} \left( 3 \kappa {\cal H} \varphi'^2 - 6 {\cal K} {\cal H} - a^2 \kappa \varphi' (\partial V / \partial \varphi) \right)}
\nonumber \\
&~~~
\times \left[
\frac{1}{2} \kappa {\cal H}^2 \Pi_\Psi^2
+ \frac{1}{2} \kappa \varphi'^2 \Pi_\Phi^2
- \kappa {\cal H} \varphi' \Pi_\Phi \Pi_\Psi
\right.
\nonumber \\
&~
\left.
~
+ (6 a^2 \sqrt{\gamma} {\cal K} \varphi' {\cal R} + \kappa \varphi' C_2) \Pi_\Phi
+
\left(
a^2 \sqrt{\gamma} \kappa (a^2 \varphi' (\partial V / \partial \varphi) - 3{\cal H} \varphi'^2) {\cal R}
- \kappa {\cal H} C_2
\right) \Pi_\Psi
\right.
\nonumber \\
&~
\left.
~
+ \frac{1}{2} \kappa C_2^2
+
\frac{a^2 \sqrt{\gamma} \left[ \left( 3 {\cal H}^2 - {\cal K} \right) \varphi'^2 - a^2 {\cal H} \varphi' (\partial V / \partial \varphi) \right]}{Q} (3 a^2 \sqrt{\gamma} {\cal K} {\cal R}^2 + \kappa {\cal R} C_2)
\right]
- \frac{1}{2} f C_2^2,
\end{align}
we can reproduce the transformations via
\begin{align}
\Psi
&= - \frac{\partial W}{\partial \Pi_\Psi},
~~~~~~
\Phi
= - \frac{\partial W}{\partial \Pi_\Phi},
~~~~~~
\Pi_{\cal R}
= - \frac{\partial W}{\partial {\cal R}},
~~~~~~
\Pi_{C_2}
= - \frac{\partial W}{\partial C_2}.
\end{align}
The new (total) Hamiltonian becomes
\begin{align}
K_T
&=
H_T + \left. \frac{\partial W}{\partial \tau} \right|_{\Pi_\Psi = \Pi_\Psi ({\cal R}, C_2, \Pi_{\cal R}, \Pi_{C_2}),~\Pi_\Phi = \Pi_\Phi ({\cal R}, C_2, \Pi_{\cal R}, \Pi_{C_2})}
\nonumber \\
&=
\frac{1}{2} \frac{Q}{a^2 \sqrt{\gamma} \varphi'^2} \Pi_{\cal R}^2
+ a^2 \sqrt{\gamma} {\cal K} \left(
- \frac{\varphi'^2}{Q} + \frac{{\cal K} \varphi'^2 + a^2 {\cal H} \varphi' (\partial V/\partial\varphi)}{Q^2} 
\right) {\cal R}^2 
\nonumber \\
&~~~
+ {\cal H} \Pi_{C_2} C_2
- \left( \frac{\kappa}{12 a^2 \sqrt{\gamma} Q} + f {\cal H} + \frac{1}{2} f' \right) C_2^2 
+ \left(
- \frac{\kappa \varphi'^2}{2 Q}
+ \frac{\kappa}{3} \frac{{\cal K} \varphi'^2 + a^2 {\cal H} \varphi' (\partial V / \partial \varphi)}{Q^2} 
\right) {\cal R} C_2
\nonumber \\
&~~~
- A C_2
+ v_1 \Pi_A.
\end{align}
The Hamiltonian $K_T$ shows that the ${\cal R}$ sector and the $C_2$-$A$ sector are coupled only through the ${\cal R} C_2$ term. 
Indeed, the Hamilton equations are
\begin{align}
    {\cal R}' = \frac{\partial K_T}{\partial \Pi_{\cal R}} =& \frac{ Q}{a^2 \sqrt{\gamma} \varphi'^2 } \Pi_{\cal R}, \\
         C_2' = \frac{\partial K_T}{\partial \Pi_{C_2}}    =& {\cal H} C_2, \\
           A' = \frac{\partial K_T}{\partial \Pi_{A}}      =& v_1, \\
\Pi_{\cal R}' = -\frac{\partial K_T}{\partial {\cal R}}    =& -2 {\cal K} a^2 \sqrt{\gamma} \left( -\frac{\varphi'^2}{ Q} + \frac{ {\cal K}\varphi'^2 + a^2 {\cal H} \varphi' (\partial V/\partial\varphi) }{ Q^2 }  \right) {\cal R}, \\
   \Pi_{C_2}' = -\frac{\partial K_T}{\partial C_2}         =& -{\cal H} \Pi_{C_2} + \left( \frac{\kappa}{6a^2 \sqrt{\gamma}} \frac{1}{Q} + 2 f {\cal H} + f'
\right) C_2 \nonumber\\
	& - \left( -\frac{\kappa}{2} \frac{\varphi'^2}{Q} + \frac{\kappa}{3} \frac{ {\cal K}\varphi'^2 + a^2 {\cal H} \varphi' (\partial V/\partial\varphi)  }{Q^2}  \right) {\cal R} + A, \\
       \Pi_A' = -\frac{\partial K_T}{\partial A}           =& C_2.
\end{align}
As long as the constraints are satisfied ($\Pi_A = C_2 = 0$, which we choose as the boundary condition), the Hamilton equations for ${\cal R}$ and $\Pi_{\cal R}$ are not affected.
Thus, we can define the following reduced Hamiltonian
\begin{align}
H_R =
& \frac{1}{2} \frac{Q}{a^2 \sqrt{\gamma} \varphi'^2 } \Pi_{\cal R}^2
+ \sqrt{\gamma} {\cal K} a^2 \left( -\frac{\varphi'^2}{Q} + \frac{ {\cal K}\varphi'^2 + a^2 {\cal H} \varphi' (\partial V/\partial\varphi)}{Q^2}  \right) {\cal R}^2, \label{eq:reduced Hamiltonian}
\end{align}
with the subscript $R$ denoting ``reduced".
The above Hamiltonian shows that the sign of the kinetic term is determined by the sign of $Q$,
and thus we obtain the same condition for the negative mode condition as Sec.~\ref{sec:lagrangian formulation}.
This Hamiltonian is the same as the one directly constructed from the gauge invariant Lagrangian (\ref{eq:L_gaugeinv}).

\subsection{Dirac bracket approach} \label{sec:dirac bracket}

Let us take another approach which utilizes the Dirac bracket \cite{Dirac}.
We impose a gauge fixing condition $\chi_1 = 0$ such that $[\chi_1, C_1] = 0$ and $[\chi_1, C_2] \neq 0$ are satisfied \cite{Sugano:1982bm}.
For consistency, we have to impose $\chi_2 \equiv \partial\chi_1/\partial\tau + [\chi_1, H_T] = 0$.
No more condition appears because
$\partial\chi_2/\partial\tau + [\chi_2, H_T] = 0$ can be maintained by choosing an appropriate $v_1(\tau)$ in $H_T$.
The Dirac bracket is defined as
\begin{align}
[\alpha,\beta]_D \equiv [\alpha,\beta] - \sum_{i,j=1}^4 [\alpha,\phi_i] M^{-1}_{ij} [\phi_j,\beta], \qquad M_{ij} = [\phi_i, \phi_j], \label{eq:dirac bracket}
\end{align}
where $(\phi_1, \phi_2, \phi_3, \phi_4) = (C_1, C_2, \chi_1, \chi_2)$.
The time evolution of an arbitrary variable $\alpha$ (see, \textit{e.g.}, Ref.~\cite{de1996time}) is
\begin{align}
\alpha' = \frac{\partial \alpha}{\partial \tau} - \sum_{i,j=1}^4 \frac{\partial\phi_i}{\partial\tau} M^{-1}_{ij} [\phi_j,\alpha] + [\alpha,H_T]_D. \label{eq:time derivative}
\end{align}

As we have discussed so far, an infinite number of negative modes appear when the kinetic term has the wrong sign.
Let us briefly discuss how we can extract the coefficient of the kinetic term, using the following Hamiltonian with time-dependent coefficients $m(t)$, $\omega(t)$, and $a(t)$:
\begin{align}
H = \frac{p^2}{2m} + \frac{1}{2} m\omega^2 q^2 + a p q.
\end{align}
The coefficient of the kinetic term can be extracted by the Poisson bracket $[q, dq/dt]$ for the coordinate $q$:
\begin{align}
\left[q, \frac{dq}{dt} \right] = \left[q, \frac{p}{m} + a q\right] = \frac{1}{m}.
\end{align}
Note that this is essentially the inverse of the coefficient of $\dot{q}^2$ when we Legendre-transform back to the Lagrangian formulation:
\begin{align}
L
&= \left[ p \dot{q} - H \right]_{p = p(q, \dot{q})} 
= \frac{m}{2} \dot{q}^2 - m a q \dot{q} + \frac{m}{2} (a^2 - \omega^2) q^2,
\end{align}
and thus we can discuss the negative mode condition using $[q, dq/dt]$.

Note that the coefficient of the kinetic term changes if we use a different coordinate.
Indeed, the Poisson bracket for a general linear combination $Q \equiv c_q q+c_p p$ is given as
\begin{align}
\left[Q, \frac{dQ}{dt} \right] 
&= 
\left[Q, \frac{\partial Q}{\partial t} + [Q,H] \right] 
=
\frac{c_q^2}{m} + c_p^2 m\omega^2 + c_q \frac{dc_p}{dt}  - \frac{dc_q}{dt} c_p. \label{eq:new kinetic term}
\end{align}
This indicates that, once we canonical transform from $(p,q)$ to $(P,Q)$ and then Legendre-transform back to the Lagrangian formulation, the coefficient of the kinetic term takes a different form.
Thus, the choice of the coordinate $q$ is essential when we discuss the sign of the kinetic term.
The same argument also holds for the Dirac bracket.
Therefore we can calculate the coefficient of the kinetic term for the fluctuation using the Dirac bracket $[q, q']_D$ once we specify the fluctuation $q$.
We discuss explicit examples related to this point in the following subsections and in Appendix~\ref{sec:comparison}.

\subsubsection{Comparison with LRT \cite{Lavrelashvili:1985vn}, LW \cite{Lee:2014uza}, and L-I \cite{Lavrelashvili:1999sr}}
\label{sec:comparison with LRT}

Inspired by the Lagrangian approach, we discuss the kinetic term of $\xi$ defined as
\begin{align}
\xi \equiv g(\tau)\left( \varphi' \Psi + {\cal H} \Phi \right),
\end{align}
where $g(\tau)$ is an arbitrary function of $\tau$.
Here we take a generic gauge fixing condition
\begin{align}
\chi_1 = c_\Phi(\tau) \Phi + c_\Psi(\tau) \Psi + c_{\Pi_\Phi}(\tau) \Pi_\Phi + c_{\Pi_\Psi}(\tau) \Pi_\Psi.
\end{align}
For consistency we also need to impose $\chi_2 \equiv \partial \chi_1/\partial\tau + [\chi_1, H_T] = 0$.
$\chi_2$ is given as
\begin{align}
\chi_2
=&  c_\Phi \left( \frac{\Pi_\Phi}{a^2\sqrt{\gamma}} + \varphi' A  \right) 
   +c_\Psi \left( -\frac{\kappa}{6a^2\sqrt{\gamma}} \Pi_\Psi + \frac{1}{2}\kappa \varphi' \Phi - {\cal H} A \right) \nonumber\\
 & +c_{\Pi_\Phi} \left[ -\frac{1}{2}\kappa\varphi'\Pi_\Psi + a^2 \sqrt{\gamma}\left( a^2 \frac{\partial^2 V}{\partial\varphi^2} + \frac{3}{2}\kappa \varphi'^2 \right)\Phi - a^2 \sqrt{\gamma} \left( 3\varphi' {\cal H} - a^2 \frac{\partial V}{\partial\varphi} \right) A \right] \nonumber\\
 & +c_{\Pi_\Psi} \frac{6{\cal K}}{\kappa} a^2 \sqrt{\gamma} (\Psi -A)  
   +c_\Phi' \Phi + c_\Psi' \Psi + c_{\Pi_\Phi}' \Pi_\Phi + c_{\Pi_\Psi}' \Pi_\Psi. \label{eq:chi2}
\end{align}
By using Eq.~(\ref{eq:dirac bracket}) and Eq.~(\ref{eq:time derivative}), a straightforward calculation gives
\begin{align}
[\xi, \xi']_D
= \frac{ g^2 Q }{a^2 \sqrt{\gamma}}. \label{eq:dirac bracket xi}
\end{align}
An infinite number of negative modes of $\xi$ appear if the above expression is negative,
and thus we obtain the same condition for the negative mode condition as Sec.~\ref{sec:lagrangian formulation},
LRT \cite{Lavrelashvili:1985vn}, LW \cite{Lee:2014uza}, and L-I \cite{Lavrelashvili:1999sr}.

\subsubsection{Comparison with KLT \cite{Khvedelidze:2000cp}, GT \cite{Gratton:2000fj}, and L-III \cite{Lavrelashvili:1999sr}}

Next we discuss the result by KLT \cite{Khvedelidze:2000cp} and GT \cite{Gratton:2000fj}.
Both papers discuss the kinetic term of $\Phi$ in $\Pi_\Psi = 0$ gauge.
Thus, we take $\chi_1 = \Pi_\Psi$ and impose $\chi_2 = A-\Psi$ for consistency.
Again, by using Eq.~(\ref{eq:dirac bracket}) and Eq.~(\ref{eq:time derivative}), we obtain
\begin{align}
[\Phi, \Phi' ]_D &= \frac{1}{a^2 \sqrt{\gamma}} \left( 1 - \frac{\kappa\varphi'^2}{6{\cal K}}\right). \label{eq:KLT coefficient}
\end{align}
This result is consistent with the result obtained in KLT and GT (see also L-III \cite{Lavrelashvili:1999sr}).

The above discussion relies on the specific choice of the gauge fixing condition.
Let us discuss this result in a general gauge.
The gauge invariant combination equivalent to $\Phi$ in $\Pi_\Psi = 0$ gauge is
\begin{align}
\hat\Phi \equiv \Phi + \frac{\kappa \varphi'}{6 \sqrt{\gamma} a^2 {\cal K}} \Pi_\Psi.
\end{align}
For a generic gauge condition $\chi_1 = c_\Phi(\tau) \Phi + c_\Psi(\tau) \Psi + c_{\Pi_\Phi}(\tau) \Pi_\Phi + c_{\Pi_\Psi}(\tau) \Pi_\Psi$
with $\chi_2$ given in Eq.~(\ref{eq:chi2}),
a straightforward calculation shows
\begin{align}
[\hat\Phi, \hat\Phi']_D = \frac{1}{a^2 \sqrt{\gamma}} \left( 1 - \frac{\kappa\varphi'^2}{6{\cal K}}\right). \label{eq:dirac bracket phi}
\end{align}
The sign of the kinetic term of $\xi$ and $\hat\Phi$ do not coincide in general.
We have three independent gauge invariant variable; ${\cal R}$, $\Pi_{\cal R}$, and $C_2$.
Thus, $\hat\Phi$ can be written as a linear combination of them.
By using Eq.~(\ref{eq:definition R}), Eq.~(\ref{eq:Rprime}), Eq.~(\ref{eq:definition PiR}), and Eq.~(\ref{eq:C1 time evolution}),
it is straightforward to show the following relation:
\begin{align}
\hat\Phi =
	\frac{ {\cal H}\varphi' }{Q} {\cal R}
	+ \frac{\kappa\varphi'}{6\sqrt{\gamma} a^2 {\cal K}} \Pi_{\cal R}
	+ \frac{\kappa\varphi' {\cal H}}{6\sqrt{\gamma} a^2 {\cal K} Q} C_2.
\end{align}
Also, $\xi$ discussed in Sec.~\ref{sec:comparison with LRT} is proportional to ${\cal R}$.
Thus, $\Phi$ in KLT can be understood as a mixture of gauge invariant fluctuation and its conjugate momentum in LRT and LW.
This explains the reason of difference between $[\xi, \xi']_D$ given in Eq.~(\ref{eq:dirac bracket xi})
 and $[\hat\Phi, \hat\Phi']_D$ given in Eq.~(\ref{eq:dirac bracket phi}).
As we have discussed in Eq.~(\ref{eq:new kinetic term}), a canonical coordinate transformation which mixes a coordinate and its conjugate momentum changes the coefficient of kinetic term.
We can obtain the Hamiltonian for $\hat\Phi$ and its conjugate momentum
from the Hamiltonian given in Eq.~(\ref{eq:reduced Hamiltonian}) by a canonical coordinate transformation from ${\cal R}$ and $\Pi_{\cal R}$. 
We can see the sign of the coefficient of the kinetic term in new coordinate is same as the sign of Eq.~(\ref{eq:KLT coefficient}).
This is the reason why the negative mode condition in KLT is different from LRT and LW.

\subsubsection{Comparison with GMST \cite{Garriga_1998,Garriga_1999} and L-II \cite{Lavrelashvili:1999sr}}

Finally we discuss L-II \cite{Lavrelashvili:1999sr},
which relies on the results given in Appendix B of GMST \cite{Garriga_1998,Garriga_1999} though GMST itself does not discuss the negative mode problem directly.
The negative mode condition which can be read off from this literature can be understood as the kinetic term of $\Psi$ in $\Phi = 0$ gauge (though the authors use gauge invariant counterparts ${\bf \Psi}$ and ${\bf \Pi}_{\bf \Psi}$: see Appendix~\ref{sec:approach 2}).
Therefore, we take $\chi_1 = \Phi$ and $\chi_2 = \chi_1' = \Pi_\Phi/a^2\sqrt{\gamma} + \varphi' A$.
By using the Dirac bracket given in Eq.~(\ref{eq:dirac bracket}), we obtain
\begin{align}
[\Psi, \Pi_\Psi]_D &= 1,
\end{align}
and
\begin{align}
[\Psi, [\Psi, H_T]_D]_D &= \frac{ {\cal H}^2 }{a^2 \sqrt{\gamma} \varphi'^2} \left( 1 - \frac{\kappa \varphi'^2}{6 {\cal H}^2 }\right), \\
[\Psi, [\Pi_\Psi, H_T]_D]_D &= \frac{6 {\cal K} {\cal H}}{\kappa \varphi'^2}, \\
[\Pi_\Psi, [\Pi_\Psi, H_T]_D]_D &= \frac{6 a^2 \sqrt{\gamma} {\cal K}}{\kappa} \frac{6{\cal K} - \kappa \varphi'^2}{\kappa \varphi'^2},
\end{align}
thus reproducing the coefficients of their Hamiltonian (see Eq.~(\ref{eq:Hamiltonian Lavrelashvili II})).
Note that we started with the Euclidean action (\ref{eq:action}) while Appendix~B of Ref.~\cite{Garriga_1998} starts with the Lorentzian action.
Note also that here we consider the $O(4)$ invariant mode only.
The coefficient of $\Pi_\Psi^2$ can be read off as
\begin{align}
[\Psi, \Psi']_D &= \frac{ {\cal H}^2 }{a^2 \sqrt{\gamma} \varphi'^2} \left( 1 - \frac{\kappa \varphi'^2}{6 {\cal H}^2 }\right),
\end{align}
which is consistent with Eq.~(B12) of Ref.~\cite{Garriga_1998}.
The authors further transform the Hamiltonian using the variables $\tilde{p}$ and $\tilde{q}$ defined as
\begin{align}
\Psi
&=
\frac{\kappa \varphi'}{4} \tilde{q} + \frac{{\cal H}}{3a^2 \sqrt{\gamma} {\cal K} \varphi'} \tilde{p},
\label{eq:def qtilde 1}
\\
\Pi_\Psi
&=
- \frac{3a^2 \sqrt{\gamma} {\cal K} \varphi'}{2 {\cal H}} \tilde{q} + \frac{2}{\kappa \varphi'} \tilde{p}.
\label{eq:def qtilde 2}
\end{align}
The coefficient of $\tilde{p}^2$ can be read off from the Dirac bracket with the same gauge fixing condition as
\begin{align}
[\tilde{q}, \tilde{q}']_D 
&= \frac{1}{3 a^2 \sqrt{\gamma} {\cal K}}.
\end{align}
This is consistent with Eq.~(B18) of Ref.~\cite{Garriga_1998} (see also Eq.~(\ref{eq:L_qtilde})).
In this case, the sign of the kinetic term for $O(4)$ symmetric fluctuation is positive definite and negative mode does not appear.
Note that the Euclidean action given in Eq.~(12) in Ref.~\cite{Lavrelashvili:1998dt} and Eq.~(38) in L-II~\cite{Lavrelashvili:1999sr}
has the opposite sign for ${\cal K}=1$. For detail, see Appendix \ref{sec:approach 2}.
L-II takes analytic continuation as $\tilde q \to -i \tilde q$ to obtain a positive kinetic term.

\section{Summary and discussions} \label{sec:summary and discussion}

It has been known that an infinite number of negative modes can appear in the fluctuations around the CDL bounce solution,
and this negative mode problem \cite{Lavrelashvili:1985vn} obscures physical interpretation on the CDL bounce solution.
In this paper, we have discussed the negative mode condition with different approaches.

In the Lagrangian formalism discussed in Sec.~\ref{sec:lagrangian formulation},
there is a clear difference between a field and its first derivative.
We can see that ${\cal R} = \Psi + ({\cal H}/\varphi')\Phi$ is the only choice for the gauge invariant dynamical variable for $O(4)$ symmetric fluctuations,
and its kinetic term becomes negative if
\begin{align}
{\cal H}^2 - \frac{\kappa}{6} \varphi'^2 < 0, \label{eq:negative mode condition LRTLW}
\end{align}
as derived in LRT \cite{Lavrelashvili:1985vn} and LW \cite{Lee:2014uza}.

On the other hand, in the Hamiltonian formalism discussed in Sec.~\ref{sec:hamiltonian formulation}, 
there is no clear difference between a field and its conjugate momentum because we can mix them by a canonical coordinate transformation.
Hence, we need to specify a fluctuation variable to discuss the sign of its kinetic term.
If we choose ${\cal R}$, we obtain the result consistent with LRT \cite{Lavrelashvili:1985vn} and LW \cite{Lee:2014uza}.
The negative mode condition is given by Eq.~(\ref{eq:negative mode condition LRTLW}).
However, if we choose a different variable, we obtain a different negative mode condition.
By choosing $\hat\Phi = \Phi + (\kappa \varphi'/6\sqrt{\gamma}a^2{\cal K})\Pi_\Psi$,
we obtain the same result as KLT \cite{Khvedelidze:2000cp}.
The kinetic term of $\hat\Phi$ becomes negative if
\begin{align}
1 - \frac{\kappa}{6{\cal K}} \varphi'^2 < 0.
\end{align}
In GMST \cite{Garriga_1998} and L-II \cite{Lavrelashvili:1998dt},
they take $\tilde q$ (and $\tilde p$) defined in Eq.~(\ref{eq:def qtilde 1}) and Eq.~(\ref{eq:def qtilde 2}),
and discuss the sign of the kinetic term.
As shown in Ref.~\cite{Lavrelashvili:1998dt},
when we take $\tilde q$ as the fluctuation variable,
the sign of the kinetic term is positive definite \cite{Garriga_1998}.
However, the sign of the kinetic term for $O(4)$ symmetric mode and high-$\ell$ mode are opposite \cite{Garriga_1998, Lavrelashvili:1998dt},
and this requires analytic continuation such as $\tilde q \to -i \tilde q$ only for specific partial wave(s).

We cannot give a conclusive answer to the question of which condition to use for the negative mode problem.
This unclear situation stems from the fact that we do not have a concrete formalism to treat quantum gravity.
The action of the Euclidean gravity is unbounded below \cite{Gibbons:1978ac}
and this is a serious obstacle when we discuss non-perturbative property of the Euclidean gravity.
The negative mode problem could be regarded as a part of this problem.
As discussed in the introduction of GT \cite{Gratton:2000fj} (see also Ref.~\cite{Tanaka:1992zw}),
the negative mode problem could be (partly) solved by taking a new Lagrangian obtained from the Hamiltonian after a canonical coordinate transformation.
Though the original and new Lagrangians give consistent equations of motion, their value is different at the point in the phase space which does not satisfy the equations of motion.
A flaw of this treatment is the fact that there is no clear criteria for the choice of the fluctuation variable.
In order to avoid these subtle theoretical issues, the Wheeler-DeWitt equation would be
another interesting tool to analyze vacuum decay with gravity \cite{deAlwis:2019dkc}.

\section*{Acknowledgements}

We thank Francesco Muia for fruitful discussions.
The work of RJ is supported by Grants-in-Aid for JSPS Overseas Research Fellow (No. 201960698).
This work was supported by the Deutsche Forschungsgemeinschaft under Germany's Excellence Strategy - EXC 2121 ``Quantum Universe'' - 390833306.

\appendix

\section{Comparison with the literature}
\label{sec:comparison}

In this appendix, we review results derived in the literature.
We also give a comparison with the inflationary quadratic action.
In the main text, we discussed the $O(4)$ symmetric case only, but in this Appendix \ref{sec:inflation},
we do not assume $O(4)$ symmetry of fluctuation in order to compare its action with inflationary quadratic action.

\subsection{Comparison with inflationary quadratic action} \label{sec:inflation}

Though our main interest is the negative modes around the bounce solution, we can cross-check our action with the single-field inflationary quadratic action since both actions involve one scalar field and Einstein-Hilbert gravity.
We expand the action
\begin{align}
S
&= \int d^4 x \sqrt{\mp g} \left[ -\frac{1}{2\kappa} R + \frac{1}{2} (\partial_\mu \phi) (\partial^\mu\phi) \mp V(\phi) \right],
\end{align}
where the upper (lower) signs are for Lorentzian (Euclidean) metric with respect to the following perturbations
\begin{align}
ds^2
&= 
a^2 \left[
\mp (1 + 2A) d\tau^2 + 2 B_{; i} d\tau dx^i + (\gamma_{ij} (1 - 2 \Psi) + 2 E_{; ij}) dx^i dx^j
\right],
\\
\phi 
&= \varphi + \Phi.
\end{align}
Here the semicolon denoting the covariant derivative with respect to a spatial metric $\gamma_{ij}$ with a constant curvature.
We assume $\gamma_{ij}$ to be positive definite.
The resulting quadratic action is~\cite{Garriga_1998}
\begin{align}
S^{(2)} 
&= \int d\tau d^3x~{\cal L}^{(2)},
\end{align}
with
\begin{align}
{\cal L}^{(2)} 
= 
\frac{a^2 \sqrt{\gamma}}{2\kappa} \Biggl[
&- 6\Psi'^2 \mp 6{\cal K} \Psi^2 \mp 2\Psi \Delta \Psi
+ \kappa \left( \Phi'^2 \mp a^2 \frac{\partial^2 V}{\partial \varphi^2} \Phi^2 + 6\varphi' \Psi' \Phi \pm \Phi \Delta \Phi \right)
\nonumber\\
&- \left( 
12{\cal H} \Psi' \mp 12{\cal K} \Psi + 2\kappa \varphi' \Phi' \pm 2 \kappa a^2 \frac{\partial V}{\partial \varphi} \Phi \mp 4\Delta \Psi
\right) A
\nonumber\\
&- \left( 
2{\cal H}' + 4{\cal H}^2 \mp 2{\cal K}
\right) A^2
\nonumber\\
&+ \Delta (B - E') (2\kappa \varphi' \Phi - 4 \Psi') + 2{\cal K} (B - E') \Delta (B - E')
- 4{\cal H} A \Delta (B - E')
\Biggr].
\label{eq:L_full}
\end{align}
The background equations of motion are given by
\begin{align}
{\cal H}^2 - {\cal H}' \pm {\cal K} 
&= \frac{\kappa}{2}\varphi'^2, \\
2{\cal H}' + {\cal H}^2 \pm {\cal K} 
&= \frac{\kappa}{2}(- \varphi'^2 \pm 2a^2 V), \\
\varphi'' + 2{\cal H}\varphi' \pm a^2 \frac{\partial V}{\partial\varphi} 
&= 0.
\end{align}
Below we eliminate the auxiliary fields $A$ and $B - E'$.
However, before doing so, we rewrite the action in a gauge-invariant form:
\begin{align}
{\cal L}^{(2)}
&= 
a^2 \sqrt{\gamma}
\Biggl[
\frac{\varphi'^2}{2{\cal H}^2} {\cal R}'^2
+ \frac{1}{2{\cal H}^2}
\left(
- \frac{6{\cal K}^2}{\kappa} \pm (3{\cal K} + \Delta) \varphi'^2
\right)
{\cal R}^2
\nonumber \\
&~~~~~~~~~~~~
- \frac{\varphi'^2}{{\cal H}} {\cal R}' {\cal A}
\pm \frac{6{\cal K}}{\kappa} {\cal R} {\cal A}
\pm \frac{2{\cal K} \Delta}{\kappa {\cal H}} {\cal R} {\cal B}
- \frac{3 {\cal H}^2}{\kappa} {\cal A}^2 + \frac{\varphi'^2}{2} {\cal A}^2
- \frac{2 {\cal H} \Delta}{\kappa} {\cal A} {\cal B} + \frac{{\cal K} \Delta}{\kappa} {\cal B}^2
\Biggr].
\label{eq:LRAB}
\end{align}
Here ${\cal R}$, ${\cal A}$ and ${\cal B}$ are the following gauge-invariant combinations:
\begin{align}
{\cal R}
&= 
\Psi + \frac{{\cal H}}{\varphi'} \Phi,
\\
{\cal A}
&= 
A + \frac{1}{{\cal H}} \Psi' + \left( \frac{{\cal H}'}{{\cal H} \varphi'} - \frac{{\cal H}}{\varphi'} \right) \Phi,
\\
{\cal B}
&= 
B - E' \mp \frac{1}{{\cal H}} \Psi.
\end{align}
Also, in the above Lagrangian we wrote $\Delta {\cal R}^2 = {\cal R} \Delta {\cal R}$, $\Delta {\cal A} {\cal B} = {\cal A} \Delta {\cal B}$ and so on for notational simplicity.
We can integrate out ${\cal A}$ and ${\cal B}$ from the Lagrangian (\ref{eq:LRAB}).
The minimization condition gives
\begin{align}
{\cal A}
&= 
\frac{\pm (6 {\cal K}^2 {\cal H} + 2 {\cal K} {\cal H} \Delta) {\cal R} - \kappa {\cal K} \Phi'^2 {\cal R}'}
{{\cal H} (2 (3{\cal K} + \Delta) {\cal H}^2 - \kappa {\cal K} \Phi'^2)},
\label{eq:A_minimize}
\\
{\cal B}
&= 
\frac{\pm \kappa {\cal K} \Phi'^2 {\cal R} - \kappa {\cal H} \Phi'^2 {\cal R}'}
{{\cal H} (2 (3{\cal K} + \Delta) {\cal H}^2 - \kappa {\cal K} \Phi'^2)},
\label{eq:B_minimize}
\end{align}
and the resulting Lagrangian becomes
\small
\begin{align}
{\cal L}^{(2)}
&= 
a^2 \sqrt{\gamma}
\Biggl[
\frac{(3{\cal K} + \Delta) \kappa \varphi'^2 {\cal R}'^2}{2(3{\cal K} + \Delta) {\cal H}^2 - \kappa {\cal K} \varphi'^2}
+ \frac{(3{\cal K} + \Delta){\cal R}^2}{(2(3{\cal K} + \Delta) {\cal H}^2 - \kappa {\cal K} \varphi'^2)^2}
\nonumber \\
&~~~~~~~~
\times
\left(
(3{\cal K} + \Delta) 
\left(
\mp 2 {\cal H}^2 \varphi'^2 (2{\cal K} - \Delta)
- 4{\cal K}^2 \varphi'^2 
- 4{\cal K} {\cal H} \varphi' a^2 \frac{\partial V}{\partial \varphi}
\right)
\pm
\kappa {\cal K} \varphi'^4 (2{\cal K} + \Delta)
\right)
\Biggr].
\end{align}
\normalsize
The same action can be obtained if we directly integrate out $A$ and $B - E'$ from the action (\ref{eq:L_full}).
This action reduces to the single-field inflationary quadratic action in the zero curvature limit ${\cal K} \to 0$ and $\gamma_{ij} \to \delta_{ij}$(see {\it e.g.} Ref.~\cite{Maldacena:2002vr}):
\begin{align}
{\cal L}^{(2)}
&\xrightarrow{{\cal K} \to 0}
\frac{a^2 \varphi'^2}{2{\cal H}^2}
\left(
{\cal R}'^2 \pm {\cal R} \Delta {\cal R}
\right).
\end{align}
The other limit is the zero mode $\Delta \to 0$. We get 
\small
\begin{align}
{\cal L}^{(2)}
&\xrightarrow{\Delta \to 0}
a^2 \sqrt{\gamma}
\Biggl[
\frac{3 \varphi'^2}{6{\cal H}^2 - \kappa \varphi'^2} {\cal R}'^2
+ 
\frac{6 {\cal K}}{(6{\cal H}^2 - \kappa \varphi'^2)^2}
\left(
\mp 6 {\cal H}^2 \varphi'^2 - 6 {\cal K} \varphi'^2 \pm \kappa \varphi'^4 - 6 {\cal H} \varphi' \frac{\partial V}{\partial \varphi}
\right)
{\cal R}^2
\Biggr].
\end{align}
\normalsize
In Sec.~\ref{sec:lagrangian formulation} we discussed the negative mode condition based on this Lagrangian.

\subsection{Comparison with L-I \cite{Lavrelashvili:1999sr}}

The authors start with the Lagrangian~(\ref{eq:quadratic}). They take the gauge
\begin{align}
\Psi
&= 0,
\end{align}
and eliminate $A$ as an auxiliary field.
This gauge condition corresponds to taking $\epsilon = a \Psi / \varphi'$ in Eq.~(\ref{eq:gauge_transformation}).
The resulting action is\footnote{
Note that the square for the term $(\partial V / \partial \varphi)^2$ is missing in Eq.~(33) of Ref.~\cite{Lavrelashvili:1999sr}.
}
\begin{align}
S^{(2)} 
&= \int d\tau d^3x~
\frac{a^2 \sqrt{\gamma}}{2\kappa} 
\nonumber \\
&~~~~
\times \left[
\kappa \left(\Phi'^2 + a^2 \frac{\partial^2 V}{\partial \varphi^2} \Phi^2 \right)
- \left( 2\kappa \varphi' \Phi' - 2\kappa a^2 \frac{\partial V}{\partial \varphi} \Phi  \right) A
- 2({\cal H}' + 2{\cal H}^2 + {\cal K}) A^2
\right]
\nonumber \\
&= \int d\tau d^3x~
\frac{a^2 \sqrt{\gamma} {\cal H}^2}{2Q} 
\left[
\Phi'^2
- \kappa a^2 \frac{\partial V}{\partial \varphi} \frac{\varphi'}{ 3  {\cal H}^2} \Phi' \Phi
+ \left( \frac{ \kappa }{6{\cal H}^2} \left( a^2 \frac{\partial V}{\partial \varphi}\right)^2 
+ a^2 \frac{Q}{{\cal H}^2} \frac{\partial^2 V}{\partial \varphi^2} \right) \Phi^2
\right].
\end{align}
They read off the negative mode condition as
\begin{align}
Q
&= 
{\cal H}^2 - \frac{\kappa \varphi'^2}{6}.
\label{eq:Qapp1}
\end{align}

\subsection{Comparison with GMST \cite{Garriga_1998,Garriga_1999} and L-II \cite{Lavrelashvili:1999sr}}\label{sec:approach 2}

We follow Appendix~B of Ref.~\cite{Garriga_1998}, focusing on the $O(4)$ invariant mode only.
Approach II of Ref.~\cite{Lavrelashvili:1999sr} uses equations derived in Ref.~\cite{Garriga_1998}.
Ref.~\cite{Garriga_1998} starts from the Lorentzian action while we discussed the Euclidean action in Sec.~\ref{sec:hamiltonian formulation}.
Here we keep both conventions.
We start with the Lagrangian (\ref{eq:L_full}) with $\Delta \to 0$ and define canonical momenta as
\begin{align}
\Pi_\Psi
&= 
\frac{2a^2 \sqrt{\gamma}}{\kappa}
\left(
- 3\Psi' + \frac{3 \kappa}{2} \varphi' \Phi - 3{\cal H} A 
\right),
\\
\Pi_\Phi
&= 
a^2 \sqrt{\gamma}
\left(
\Phi' - \varphi' A
\right).
\end{align}
The constrained Hamiltonian $H_C$ is constructed from
\begin{align}
{\cal L}^{(2)}
&= 
\Pi_\Psi \Psi' + 
\Pi_\Phi \Phi'
- H_C,
\end{align}
and it becomes
\begin{align}
H_C
&=
\frac{\kappa}{12 a^2 \sqrt{\gamma} {\cal K}}
\left(
- {\cal K} \Pi_\Psi^2 + \frac{6 {\cal K}}{\kappa} \Pi_\Phi^2
\right)
+ \frac{\kappa \varphi'}{2} \Pi_\Psi \Phi
\nonumber \\
&~~~~
+ a^2 \sqrt{\gamma}
\left[
\pm \frac{3 {\cal K}}{\kappa} \Psi^2 
+ \frac{1}{2} 
\left(
\mp 3 {\cal K} + {\cal H}^2 + {\cal H}' - \frac{\varphi'''}{\varphi'}
\right) \Phi^2
\right]
- A C_2.
\end{align}
Here the upper (lower) signs are for Lorentzian (Euclidean) metric, and $C_2$ is given in Eq.~(\ref{eq:C1 time evolution})
\begin{align}
C_2 
&= 
- \varphi' \Pi_\Phi + {\cal H} \Pi_\Psi + a^2 \sqrt{\gamma} \left[ \pm \frac{6{\cal K}}{\kappa} \Psi + \left( \varphi'' - {\cal H} \varphi' \right) \Phi \right]
\nonumber \\
&= 
- \varphi' \Pi_\Phi + {\cal H} \Pi_\Psi + a^2 \sqrt{\gamma} \left[ \pm \frac{6{\cal K}}{\kappa}\Psi + \left( - 3{\cal H}\varphi' \mp a^2 \frac{\partial V}{\partial\varphi} \right) \Phi \right].
\end{align}
By using $C_2$, ${\cal L}^{(2)}$ can be rewritten as
\begin{align}
{\cal L}^{(2)}[\Pi_\Phi, \Pi_\Psi, \Psi, \Phi] &= {\cal L}^{(2)*}[\Pi_\Psi, \Psi, \Phi] + \frac{C_2^2}{2 a^2 \sqrt{\gamma} \varphi'^2}.
\end{align}
Since the RHS does not contain any linear terms in $C_2$,
the equations of motion from ${\cal L}^{(2)*}$ is consistent with those from ${\cal L}^{(2)}$ as long as $C_2 = 0$ is satisfied.
Thus, we can eliminate $\Pi_\Phi$ by using $C_2 = 0$ and use ${\cal L}^{(2)*}$ as the Lagrangian.
The Lagrangian ${\cal L}^{(2)*}$ can be simplified by adopting gauge invariant variables
\begin{align}
{\bf \Psi}
&\equiv
\Psi + \frac{{\cal H}}{\varphi'} \Phi,
\\
{\bf \Pi}_{\bf \Psi}
&\equiv
\Pi_\Psi \mp \frac{6 a^2 \sqrt{\gamma} {\cal K}}{\kappa \varphi'} \Phi.
\end{align}
In Sec.~\ref{sec:dirac bracket} we fixed the gauge to be $\Phi = 0$ rather than using these gauge invariant variables ${\bf \Psi}$ and ${\bf \Pi}_{\bf \Psi}$.
Here we proceed with ${\bf \Psi}$ and ${\bf \Pi}_{\bf \Psi}$.
By using ${\bf \Psi}$ and ${\bf \Pi}_{\bf \Psi}$, the reduced Lagrangian (denoted by the superscript ``$*$'') becomes
\begin{align}
{\cal L}^{(2)*} &= {\bf \Pi}_{\bf \Psi} {\bf \Psi}' - H_C^*,
\end{align}
where the constrained Hamiltonian is
\begin{align}
H_C^*
&= 
\frac{2 a^2 \sqrt{\gamma}}{\varphi'^2}
\left(
\pm \frac{3 {\cal K}}{\kappa} {\bf \Psi} + \frac{{\cal H}}{2 a^2 \sqrt{\gamma}} {\bf \Pi}_{\bf \Psi}
\right)^2
\pm \frac{3a^2 \sqrt{\gamma}}{\kappa} {\cal K} {\bf \Psi}^2
- \frac{\kappa}{12 a^2 \sqrt{\gamma}} {\bf \Pi}_{\bf \Psi}^2
\nonumber \\
&= 
a^2 \sqrt{\gamma} 
\left[
\frac{6 {\cal H}^2 - \kappa \varphi'^2}{12 \varphi'^2} \left( \frac{{\bf \Pi}_{\bf \Psi}}{a^2 \sqrt{\gamma}} \right)^2
\pm \frac{6 {\cal K} {\cal H}}{\kappa \varphi'^2} \frac{{\bf \Pi}_{\bf \Psi}}{a^2 \sqrt{\gamma}} {\bf \Psi}
+ \frac{3 {\cal K}}{\kappa} \frac{6 {\cal K} \pm \kappa \varphi'^2}{\kappa \varphi'^2} {\bf \Psi}^2
\right]. 
\label{eq:Hamiltonian Lavrelashvili II}
\end{align}
We next perform a canonical transformation by a (type-three) generating function
\begin{align}
W
&=
\mp \frac{3 a^2 \sqrt{\gamma} {\cal K}}{{\cal H}}
\left(
\frac{1}{\kappa} {\bf \Psi}^2
- \varphi' {\bf \Psi} {\bf \tilde{q}}
+ \frac{\kappa \varphi'^2}{8} {\bf \tilde{q}}^2
\right).
\end{align}
The relation between the original variables $({\bf \Psi}, {\bf \Pi}_{\bf \Psi})$ and new ones $({\bf \tilde{q}}, {\bf \tilde{p}})$  is obtained from ${\bf \Pi}_{\bf \Psi} = \partial W / \partial {\bf \Psi}$ and ${\bf \tilde{p}} = - \partial W / \partial {\bf \tilde{q}}$, which gives\footnote{
As discussed in Eq.~(B20) of Ref.~\cite{Garriga_1998}, the variable ${\bf \tilde{q}}$ is equivalent to the Bardeen potential after substituting the minimization condition of the auxiliary fields.
}:
\begin{align}
{\bf \Psi}
&=
\frac{\kappa \varphi'}{4} {\bf \tilde{q}} \mp \frac{{\cal H}}{3 a^2 \sqrt{\gamma} {\cal K} \varphi'} {\bf \tilde{p}},
\label{eq:psi q p}\\
{\bf \Pi}_{\bf \Psi}
&=
\pm \frac{3 a^2 \sqrt{\gamma} {\cal K} \varphi'}{2 {\cal H}} {\bf \tilde{q}} + \frac{2}{\kappa \varphi'} {\bf \tilde{p}}.
\label{eq:pipsi q p}
\end{align}
The new Hamiltonian $K_C^*$ becomes
\begin{align}
K_C^*
&=
H_C^* + \frac{\partial W}{\partial \tau}
\nonumber \\
&=
\mp \frac{1}{6 a^2 \sqrt{\gamma} {\cal K}} {\bf \tilde{p}}^2
+ \left(
{\cal H} - \frac{\kappa \varphi'^2}{4 {\cal H}} + \frac{a^2}{\varphi'} \frac{dV}{d\varphi}
\right) {\bf \tilde{p}} {\bf \tilde{q}}
\nonumber \\
&~~~\,
+ a^2 \sqrt{\gamma} {\cal K} \left[
\mp \frac{9 {\cal H}^2}{2} \pm \frac{3 \kappa \varphi'^2}{2} \pm \frac{\kappa^2 \varphi'^4}{32 {\cal H}^2} 
+ \kappa a^2 \left( \frac{3}{2} - \frac{\kappa \varphi'^2}{8 {\cal H}^2} \right) V
\right] {\bf \tilde{q}}^2.
\end{align}
After a Legendre transformation, we get a new Lagrangian
\begin{align}
{\cal L}^{(2), {\rm new}}
&=
\mp \frac{3 a^2 \sqrt{\gamma} {\cal K}}{2}
\left(
{\bf \tilde{q}}'^2 \mp a^2 m^2 {\bf \tilde{q}}^2
\right),
\label{eq:L_qtilde}
\end{align}
with $m^2$ being
\begin{align}
m^2
&=
- \frac{1}{a^2}
\left[
4 {\cal K} \mp 2 {\cal H}' \pm \varphi' \left( \frac{1}{\varphi'} \right)''
\right]
\nonumber \\
&=
\mp \frac{2 \kappa \varphi'^2}{3 a^2}
- \left(
\frac{4 \kappa}{3} V
+ \frac{8 {\cal H}}{\varphi'} \frac{dV}{d\varphi}
+ \frac{d^2V}{d\varphi^2}
\right)
\mp \frac{2 a^2}{\varphi'^2} \left( \frac{dV}{d\varphi} \right)^2.
\end{align}
In Sec.~\ref{sec:dirac bracket} we discussed the coefficient $3 a^2 \sqrt{\gamma} {\cal K} / 2$ (starting from the Euclidean action) using the Dirac bracket.
Eq.~(\ref{eq:L_qtilde}) is consistent with
Eq.~(23) in Ref.~\cite{Gratton:1999ya} and
Eq.~(B18) in Ref.~\cite{Garriga_1998}.
See also Eq.~(B19) in Ref.~\cite{Garriga_1998} and its errata \cite{Garriga_1999}.

Eq.~(10) in Ref.~\cite{Lavrelashvili:1998dt} is consistent with the Lorentzian action given in Eq.~(\ref{eq:L_qtilde}) for a closed universe ${\cal K} = 1$.
On the other hand, the Euclidean action given in Eq.~(12) in Ref.~\cite{Lavrelashvili:1998dt} and Eq.~(38) in Ref.~\cite{Lavrelashvili:1999sr} has the opposite sign.
Ref.~\cite{Lavrelashvili:1998dt} derives the Euclidean action from the Lorentzian action by using analytic continuation.
As shown in Eqs.~(\ref{eq:psi q p}) and (\ref{eq:pipsi q p}), the relation between  ${\bf \tilde{q}}$ and ${\bf \Psi}$ includes $\varphi'$,
and the sign of kinetic term of ${\bf \tilde{q}}$ is flipped if we replace $\varphi'$ to $i\varphi'$.
Ref.~\cite{Lavrelashvili:1998dt} does not adopt this replacement and the signs of kinetic term of ${\bf \tilde{q}}$ are the same for both the Lorentzian and Euclidean actions.

\subsection{Comparison with L-III \cite{Lavrelashvili:1999sr} and KLT \cite{Khvedelidze:2000cp}} \label{sec:approach 3}

Again, let us start from the Hamiltonian $H_T$ given in Eq.~(\ref{eq:HT}).
In Ref.~\cite{Khvedelidze:2000cp}, the authors set two gauge fixing conditions
\begin{align}
A-\Psi = 0, \qquad \Pi_\Psi = 0,
\label{eq:gauge fixing approach III}
\end{align}
and eliminate $A$ and $\Pi_\Psi$ from the Hamiltonian.
The reduced Hamiltonian is obtained as
\begin{align}
H_{\rm KLT}
=
& \frac{1}{2a^2 \sqrt{\gamma}} \left( 1 - \frac{\kappa\varphi'^2}{6{\cal K}}\right) \Pi_\Phi^2 
  +\frac{\kappa\varphi'}{6{\cal K}}\left( a^2 \frac{\partial V}{\partial\varphi} -3{\cal H} \varphi' \right) \Pi_\Phi \Phi \nonumber\\
& -\frac{1}{2}a^2 \sqrt{\gamma}\left[ \frac{\kappa}{6{\cal K}}\left( a^2 \frac{\partial V}{\partial\varphi} - 3\varphi' {\cal H} \right)^2 + a^2 \frac{\partial^2 V}{\partial\varphi^2} + \frac{3}{2}\kappa\varphi'^2 \right] \Phi^2.
\label{eq:Hamiltonian Lavrelashvili III}
\end{align}
Ref.~\cite{Lavrelashvili:1999sr} derives the same effective Hamiltonian in approach III
even though their gauge fixing condition ($A = 0$, $\Pi_\Psi = 0$) is inconsistent with the time evolution.
Ref.~\cite{Lavrelashvili:1999sr} eliminates $\Psi$, $\Pi_\Psi$, $A$, and $\Pi_A$ by using constraints and gauge fixing conditions.
We can justify this elimination of variables as discussed in Appendix \ref{sec:elimination}.
The discussion in L-III and KLT is the same as the discussion utilizing the Dirac bracket in the main text,
and we obtain the consistent results.
For details, see Sec.~\ref{sec:dirac bracket}.

\subsection{Comparison with GT~\cite{Gratton:2000fj}} \label{sec:approach GT}

Ref.~\cite{Gratton:2000fj} starts from the action given in Eq.~(18) in Ref.~\cite{Gratton:1999ya}.
In this section we follow the notation of Refs.~\cite{Gratton:1999ya, Gratton:2000fj}, keeping signs for both Lorentzian and Euclidean setups.
We rewrite Eq.~(\ref{eq:L_full}) for $O(4)$ symmetric fluctuations in the notation of Refs.~\cite{Gratton:1999ya, Gratton:2000fj} as
\begin{align}
S^{(2)}
&=
\int d\tau d^3 x~
\frac{a^2 \sqrt{\gamma}}{2 \kappa}
\left[
- 6 \psi'^2 - 12 {\cal H} A \psi' - 2 ({\cal H}' + 2 {\cal H}^2) A^2
\right. \nonumber \\
&~~~~~~~~~~~~~~~~~~~~~~~~~
+ \kappa (\delta \phi'^2 \mp a^2 V_{,\phi\phi} \delta \phi^2 )
+ 2 \kappa (3 \phi_0' \psi' \delta \phi - \phi_0' \delta \phi' A \mp a^2 V_{,\phi} A \delta \phi) 
\nonumber \\[1.5ex]
&~~~~~~~~~~~~~~~~~~~~~~~~~
\left.
\pm {\cal K} (- 6 \psi^2 + 2 A^2 + 12 \psi A) \right],
\label{eq:action eq18 astro-ph/9902265}
\end{align}
where the upper (lower) signs are for Lorentzian (Euclidean) metric.
For $O(4)$ non-symmetric fluctuations, see Eq.~(18) in Ref.~\cite{Gratton:1999ya}.
Conjugate momenta for $\psi$ and $\delta\phi$ are introduced as
\begin{align}
\Pi_\psi 
&\equiv
\frac{2 a^2 \sqrt{\gamma}}{\kappa} \left( - 3 \psi' + \frac{3}{2} \kappa \phi_0'\delta \phi - 3 {\cal H} A \right), 
\nonumber \\
\Pi_{\delta\phi} 
&\equiv 
a^2 \sqrt{\gamma} (\delta \phi' - \phi_0' A),
\end{align}
and the action can be rewritten as
\begin{align}
S^{(2)} = 
& \int d\tau d^3 x
\left[
\Pi_\psi \psi' +
\Pi_{\delta\phi} \delta\phi'  
- \frac{\kappa}{12 a^2 \sqrt{\gamma} {\cal K}} \left( - {\cal K} \Pi_\psi^2 + \frac{6 {\cal K}}{\kappa} \Pi_{\delta \phi}^2 \right) - \frac{\kappa \phi_0'}{2} \Pi_\psi \delta \phi 
\right. \nonumber \\
&~~~~~~~~~~~~~~
\left.
- a^2 \sqrt{\gamma} \left\{ \pm \frac{3{\cal K}}{\kappa} \psi^2 + \frac{1}{2} \left( \mp 3{\cal K}
+ {\cal H}^2 + {\cal H}' - \frac{\phi'''_0}{\phi_0'} \right) \delta \phi^2 \right\} + A C_2 \right].
\end{align}
Here $C_2$ is the same constraint as Appendix~\ref{sec:approach 2}. 
In the notation of Refs.~\cite{Gratton:1999ya, Gratton:2000fj},
\begin{align}
C_2 
&\equiv
- \phi'_0 \Pi_{\delta\phi} + {\cal H} \Pi_\psi 
+ a^2 \sqrt{\gamma} \left[ \pm \frac{6{\cal K}}{\kappa} \psi + (\phi_0'' - {\cal H} \phi_0') \delta \phi \right].
\end{align}
By using $C_2$, the action can be rewritten as
\begin{align}
S^{(2)}
&=
\int d\tau d^3x~
\frac{a^2 \sqrt{\gamma}}{3 {\cal K}}
\left[
\pm \frac{2}{\kappa \phi_0'} \Psi_l \delta \phi_l' 
\pm \frac{2({\cal H}\phi_0' - \phi_0'')}{\kappa \phi_0'^2} \Psi_l \delta \phi_l
\pm \frac{1}{2} \delta \phi_l^2 
\mp \frac{1}{\kappa} \left( 1 \pm \frac{6{\cal K}}{\kappa \phi_0'^2} \right) \Psi_l^2
\right]
\nonumber \\
&\quad\quad\quad\quad\quad
+ \frac{C_2^2}{2 a^2 \sqrt{\gamma} \phi_0'^2}. \label{eq:GT Lagrangian}
\end{align}
Here $\Psi_l$ and $\delta\phi_l$ are gauge invariant variables defined as
\begin{align}
\Psi_l
&= 3 {\cal K} \psi \pm \frac{\kappa {\cal H}}{2a^2 \sqrt{\gamma}} \Pi_\psi,
\\
\delta \phi_l
&= 3 {\cal K} \delta \phi \mp \frac{\kappa \phi'}{2a^2 \sqrt{\gamma}} \Pi_\psi.
\end{align}
Since the RHS of Eq.~(\ref{eq:GT Lagrangian}) does not contain linear terms in $C_2$,
the equations of motion are not affected by dropping the $C_2^2$ term from the Lagrangian as long as $C_2 = 0$ is satisfied.
In Eq.~(\ref{eq:GT Lagrangian}), $\Pi_{\delta\phi}$ is contained only in the $C_2^2$ term.
Thus, dropping the $C_2^2$ term in Eq.~(\ref{eq:GT Lagrangian}) is equivalent to eliminating $\Pi_{\delta\phi}$ by using $C_2 = 0$:
\begin{align}
S^{(2)}
&=
\int d\tau d^3x~
\frac{a^2 \sqrt{\gamma}}{3 {\cal K}}
\left[
\pm \frac{2}{\kappa \phi_0'} \Psi_l \delta \phi_l' 
\pm \frac{2({\cal H}\phi_0' - \phi_0'')}{\kappa \phi_0'^2} \Psi_l \delta \phi_l
\pm \frac{1}{2} \delta \phi_l^2 
\mp \frac{1}{\kappa} \left( 1 \pm \frac{6{\cal K}}{\kappa \phi_0'^2} \right) \Psi_l^2
\right].
\label{eq:action eq11 hep-th/0008235}
\end{align}
After integrating $\Psi_l$ out we obtain a quadratic action for $\delta \phi_l$.
The coefficient of $\Psi_l^2$ essentially determines the negative mode condition, and we discussed it using Dirac bracket in Sec.~\ref{sec:dirac bracket}.
In the Lorentzian setup (as discussed in Refs.~\cite{Gratton:1999ya, Gratton:2000fj}) the CDL bounce corresponds to the case with ${\cal K} = 1$ (with an appropriate unit) and we should flip the sign of $\phi_0'^2$, and the negative modes appear if $1 - \kappa\phi_0'^2/6$ becomes negative.

\section{Comments on elimination of variable by constraints} \label{sec:elimination}

In KLT \cite{Khvedelidze:2000cp} and L-III \cite{Lavrelashvili:1999sr}, the Hamiltonian is reduced by eliminating variable(s) by using constraints.
For details, see Appendix \ref{sec:approach 3}.
In this Appendix we discuss the validity of this procedure.

\subsection{An example in which elimination does NOT work}

First we show that eliminating variable(s) by using constraint can give wrong equations of motion in general.
Let us take a Hamiltonian $H(q,p)$ with a constraint $C(q,p)$.
$C(q,p)=0$ can be solved for $p_n$ and let us denote the solution as
\begin{align}
p_n = f(q_1, \cdots, q_n, p_1 \cdots, p_{n-1}).
\end{align}
Then we define the reduced Hamiltonian $\tilde H$ as
\begin{align}
\tilde H(q_1, \cdots, q_n, p_1, \cdots, p_{n-1}) \equiv H(q_1, \cdots, q_n, p_1, \cdots, p_{n-1}, f(q_1, \cdots, p_{n-1})).
\end{align}
The derivatives of $\tilde H$ are given as
\begin{align}
\frac{\partial\tilde H}{\partial q_i} = \frac{\partial H}{\partial q_i}  + \frac{\partial H}{\partial p_n} \frac{\partial f}{\partial q_i}, \qquad
\frac{\partial\tilde H}{\partial p_i} = \frac{\partial H}{\partial p_i}  + \frac{\partial H}{\partial p_n} \frac{\partial f}{\partial p_i}.
\end{align}
Thus, in general, $\tilde H$ does not give the Hamilton equation which is consistent with the original Hamiltonian.

Let us see this explicitly in the following simple example.
\begin{align}
L = -\frac{3 a \dot a^2}{N} + \frac{1}{2N} a^3  \dot\phi^2 - a^3 V(\phi).
\end{align}
This is the Lagrangian for the inflaton zero mode.
The conjugate momenta for each variable are
\begin{align}
\Pi_a = -\frac{6a\dot a}{N}, \qquad
\Pi_\phi = \frac{a^3 \dot\phi}{N}, \qquad
\Pi_N = 0.
\end{align}
Thus, the Hamiltonian is
\begin{align}
H = \Pi_N \dot N + N\left( -\frac{p_a^2}{12a} + \frac{1}{2a^3}p_\phi^2 + a^3 V(\phi) \right).
\end{align}
In this system, $\Pi_N = 0$ is a primary constraint and  we obtain a secondary constraint from $\dot\Pi_N = \partial H / \partial N = 0$:
\begin{align}
-\frac{p_a^2}{12a} + \frac{1}{2a^3}p_\phi^2 + a^3 V  = 0.
\end{align}
If we simplify the Hamiltonian by using this constraint, the reduced new Hamiltonian is $\tilde H = \Pi_N \dot N$.
$\tilde H$ does not depend on $\phi$ and $a$, and apparently, $\tilde H$ does not provide the correct Hamilton equations.

\subsection{An example in which elimination works}

Next we show an example in which elimination works.
Let us discuss a system which has the following two constraints:
\begin{align}
\phi_1 = \sum_{i=1}^n (a_i q_i + b_i p_i), \qquad
\phi_2 = c q_n + d p_n. \label{eq:phi1 phi2}
\end{align}
Here we assume $a_nd-b_nc$ is non-zero.
By solving $\phi_1 = \phi_2 = 0$ for $q_n$ and $p_n$, we obtain
\begin{align}
q_n = -\frac{d}{a_nd-b_nc}\sum_{i=1}^{n-1} (a_i q_i + b_i p_i), \qquad
p_n = \frac{c}{a_nd-b_nc}\sum_{i=1}^{n-1} (a_i q_i + b_i p_i). \label{eq:qn pn solution}
\end{align}
The Dirac bracket is defined as
\begin{align}
[A,B]_D \equiv [A,B] + \frac{1}{[\phi_1,\phi_2]} \left(  [A,\phi_1] [\phi_2,B] - [A,\phi_2] [\phi_1,B] \right).
\end{align}
Let us assume that $A$ does not depend on $q_n$ and $p_n$. In this case, the Dirac bracket is
\begin{align}
[A,H]_D
=& [A,H] + \frac{1}{[\phi_1,\phi_2]} [A,\phi_1] [\phi_2,H] \nonumber\\
=& \sum_{i=1}^{n-1} \left( \frac{\partial A}{\partial q_i} \left[ \frac{\partial H}{\partial p_i} + \frac{b_i}{a_n d - b_n c} \left( -\frac{\partial H}{\partial q_n}d + \frac{\partial H}{\partial p_n}c \right) \right] \right) \nonumber\\
 & \qquad -\sum_{i=1}^{n-1} \left( \frac{\partial A}{\partial p_i} \left[ \frac{\partial H}{\partial q_i} + \frac{a_i}{a_n d - b_n c} \left( -\frac{\partial H}{\partial q_n}d + \frac{\partial H}{\partial p_n}c \right) \right] \right) \nonumber\\
=& \sum_{i=1}^{n-1} \left( \frac{\partial A}{\partial q_i} \left[ \frac{\partial H}{\partial p_i} + \frac{\partial H}{\partial q_n} \frac{\partial q_n}{\partial p_i} + \frac{\partial H}{\partial p_n} \frac{\partial p_n}{\partial p_i} \right] 
  - \frac{\partial A}{\partial p_i} \left[ \frac{\partial H}{\partial q_i} + \frac{\partial H}{\partial q_n} \frac{\partial q_n}{\partial q_i} + \frac{\partial H}{\partial p_n} \frac{\partial p_n}{\partial q_i} \right] \right).
\end{align}
Here we have used Eq.~(\ref{eq:qn pn solution}). Thus, the following relation is satisfied
\begin{align} 
[A,H]_D = [A,\tilde H].
\end{align}
where $\tilde H$ is the reduced Hamiltonian:
\begin{align}
&\tilde H (q_1, \cdots q_{n-1}, p_1, \cdots, p_{n-1}) 
\nonumber \\
&\equiv
H (q_1, \cdots q_{n-1}, q_n(q_1, \cdots, p_{n-1}) , p_1, \cdots, p_{n-1}, p_n(q_1, \cdots, p_{n-1})).
\end{align}
Therefore, we can reduce the Hamiltonian for $q_n$ and $p_n$ at least when the constraints has the form as Eq.~(\ref{eq:phi1 phi2}).
In order to derive the Hamiltonian in Eq.~(17) in Ref.~\cite{Khvedelidze:2000cp},
we have to utilize four constraints; $\Pi_\Psi = 0$, $A-\Psi=0$, $\Pi_A = 0$, and $\varphi' \Pi_\Phi - {\cal H} \Pi_\Psi + a^2 \sqrt{\gamma} [ (a^2 (\partial V/\partial\varphi) + 3\varphi' {\cal H} )\Phi - (6{\cal K}/\kappa)\Psi ] = 0$.
From above discussions, we can safely eliminate $\Psi$, $\Pi_\Psi$, and $A$ (and $\Pi_A$) from the Hamiltonian by using these four constraints.

\bibliography{ref}
\bibliographystyle{JHEP}

\end{document}